\documentclass[a4paper]{article}

\usepackage[dvips]{graphicx}
\usepackage{graphicx}
\usepackage{cite}
\usepackage{hyperref}
\usepackage[latin1]{inputenc}
\usepackage{color}

\newcommand{\dif}{{\rm d}}

\begin{document}

\title{\flushleft \bf AC loss calculation in coated conductor coils with a large number of turns\footnote{\color{blue} Accepted version of the article after referee comments. Published version as E Pardo 2013 Supercond. Sci. Technol. 105017, available at https://doi.org/10.1088/0953-2048/26/10/105017 . Copyright held by Institute of Physics (IoP).}}

\author{Enric Pardo\\
Institute of Electrical Engineering, Slovak Academy of Sciences,\\
Dubravska 9, 84104 Bratislava, Slovakia\\
enric.pardo@savba.sk}

\date{\today}

\maketitle

\begin{abstract}
For superconducting coils with many turns, it is widely believed that the magnetic field from magnetization currents is negligible compared to the one generated by the whole coil. This article shows that this is not true also for coils with a considerably large number of turns (up to thousands). We introduce a method to accurately calculate the AC loss in coils with virtually any number of turns that takes the magnetization currents into account. The current density and AC loss is analyzed for two windings made of coated conductor consisting of a stack of 32 pancake coils totaling 768 and 6400 turns, respectively.
\end{abstract}



\section{Introduction}

The AC loss in ReBCO\footnote{$Re$Ba$2$Cu$_3$0$_{7-x}$ where $Re$ is a rare earth, usually Y, Gd, Sm or a combination of those.} coated conductor coils has been a subject of intense study \cite{claassen06APL}-\cite{zhangM12APL}. However most of the works are for single pancake coils or double pancakes, with the exception of \cite{pardo12SSTb,zermeno12PhP}. This is because of the complexity of the calculations. Real windings may contain thousands of turns, such as high-field magnets \cite{trociewitz11APL}, SMES and transformers \cite{staines12SST}. Therefore, it is needed to develop calculation methods for large number of turns without degrading the accuracy.

The present state of the art is to approximate that the effect of the whole coil in a certain turn is the same as an applied magnetic field. This applied field (``background magnetic field") is computed by assuming that the current density is uniform in the rest of the turns (we name this approach as ``uniform approximation"). Afterwards, the AC loss in the turn of study is estimated by either measurements in a single tape \cite{oomen03SST,kawagoe04IES} or by numerical calculations \cite{tonsho04IES}. The problem of this approximation is that the neighboring turns shield the background magnetic field, in a similar way as in a stack of tapes \cite{pardo03PRB,grilli06PhC}. It is a question how important is this effect in coated-conductor windings consisting on stacks of pancake coils, where the shielding effect is expected to be high.

An important step forward for single pancake coils has been the continuous approximation \cite{prigozhin11SST}. This approximation substitutes the detailed coil cross-section by a continuous one with a $J_c$ corresponding to the cross-section average of the original coil. The current constrain is set by fixing the line integral in the axial direction to a certain value for every radial position. The continuous approximation is neither applicable to coils made of tapes or wires with non-negligible thickness of the superconducting core (such as Bi-2223 tapes or Bi-2212 or MgB$_2$ wires) nor for pancake coils with a large separation between turns.

In this article, we introduce a method to simplify the AC loss calculation without practically degrading the accuracy for general coils, including stacks of pancake coils with many pancakes, solenoids, and coils made of conductors with arbitrary thickness. In addition, we present the current density and the AC loss characteristics in two windings made of 32 pancake coils, with a total of 768 and 6400 turns, respectively.

The structure of the article is the following. First, we outline the general numerical method that takes into account the interaction of the magnetization currents between all the turns, which serves to monitor the accuracy of the considered approximations. Later, we introduce the simplified method. Subsequently, we discuss the calculated current distribution and AC loss for the two example coils made of coated conductor and evaluate the error committed by the considered approximations. Finally, we present our conclusions.


\section{Calculation method}

In this section, we first outline the general numerical method, which calculates the actual current distribution in all the turns of the coil, and afterwards we present a technique to speed up the calculations for large number of turns, suitable for any numerical method.

\subsection{General numerical method}

In this work, the basic numerical method to calculate the current density and the AC loss in the superconductor is the Minimum Magnetic Energy Variation method (MMEV), following the formulation described in \cite{HacIacinphase,pancaketheo,souc09SST}.

Summarizing, MMEV calculates the detailed current density, $J$, in the superconductor and assumes the critical-state model in its general form. That is, $|J|<J_c$ for $E=0$ and $|J|=J_c$ for $|E|>0$, where $J_c$ is the critical current density and $E$ is the electric field. MMEV uses variational principles in order to obtain $J$ and the magnetic field in the superconductor.
Variational principles for superconductors were firstly proposed by Bossavit \cite{bossavit94IEM}, although the most important contribution is the $J$ formulation from Prigozhin \cite{prigozhin97IES}. Later, Badia {\it et al.} proposed the $\bf H$ formulation \cite{badia02PRB}. MMEV is an alternative implementation to minimize the functional in the $\bf J$ formulation and set the current constrains (this method sets the current constrains directly, not through the electrostatic potential). Although Sanchez and Navau initially introduced MMEV for cylinders in the magnetization case \cite{sanchez01PRB}, the general formulation has been published by Pardo {\it et al.} in \cite{HacIacinphase,pancaketheo}. Early works on MMEV, such as \cite{sanchez01PRB}, implicitly assume that the current fronts penetrate monotonically in a half AC filed cycle. This occurs in some practical cases but not in general, specially for simultaneous applied field and transport current, like in coils \cite{HacIacinphase}.

The results of MMEV agree with experiments for coils \cite{souc09SST,pardo12SSTb}. Additionally, MMEV has been applied to model levitation \cite{delvalle07APL,delvalle11APL}. The method has also been extended to take into account the interaction of the superconductor with linear magnetic materials \cite{pancakeFM}. In this work, we take the cylindrical symmetry of the coils. Although MMEV is much faster than commercial Finite Element Methods (FEM) \cite{roebelcomp}, calculations for coils with a large number of turns are still lengthy.

\subsection{Approximations for large number of turns}
\label{s.methlarge}

Next, we present a model to speed up the calculations practically without degrading the accuracy that is applicable to any numerical method, not only for MMEV but also for models based on FEM. 

\begin{figure}[tbp]
\centering
\includegraphics[width=11cm]{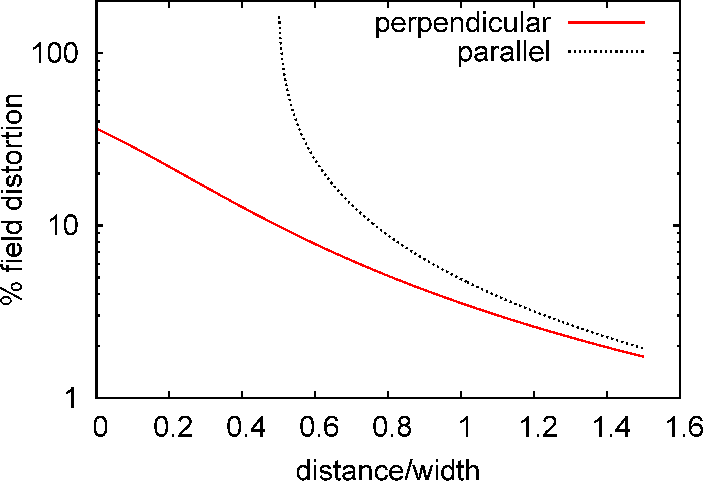}%
\caption{\label{f.distortion} The distortion of the magnetic field generated by a single tape relative to the magnetic field assuming uniform current density extends to distances of the order of the tape width. The calculations in this figure are for constant $J_c$, transport current of 20\% of the critical current and no applied magnetic field. The two curves are for lines parallel and perpendicular to the tape surface, respectively, with origin at the center of the tape (only the region outside the superconductor is shown).}
\end{figure}

In order to calculate the current distribution in a certain turn of the coil, the most straightforward approximation is to assume that the current density in all the other turns is uniform (``uniform" approximation). This approach neglects the magnetic shielding effect originated by the magnetization currents in the neighbouring turns, which often introduces significant errors. However, this is the approximation that is most commonly used in practice. In particular, the neighbour approximation calculates $J$ (or other state variable) in a certain turn $i$ by taking the field created by all the other turns as an applied field, ${\bf B}_a$, and setting the constrain that the net current is a certain given value, $I$. Since we assume uniform $J$ in all turns except $i$, the effective applied field ${\bf B}_a$ and vector potential $A_a$ (the applied vector potential is necessary for methods based on vector potential formulation, such as MMEV) are
\begin{eqnarray}
{\bf B}_a({\bf r})=I\sum_{j\neq i} \frac{1}{S_j}{\bf b}_j({\bf r}), \\
A_a({\bf r})=I\sum_{j\neq i}\frac{1}{S_j}a_j({\bf r}). \label{Aungen}
\end{eqnarray}
where $\bf r$ is the vector position, the sum is done in all the turns $j$ (except for $j=i$), $S_j$ is the cross-section area of turn $j$, and ${\bf b}_j({\bf r})$ and $a_j({\bf r})$ are the magnetic field and vector potential, respectively, created by turn $j$ per unit $J$, assuming uniform $J$. In the equation above we took Coulomb's gauge for the vector potential \cite{acreview} and used that for cylindrical symmetries, the vector potential is perpendicular to the tape cross-section, and hence only has one component. Since terms ${\bf b}_j$ and $a_j$ only depend on the shape of the turns, ${\bf B}_a$ and $A_a$ are proportional to $I$. The appendix details the numerical process to calculate $a_j$ for the computations in this article.

A more accurate approach is to assume uniform current distribution only for the turns farther than the turn of study, $i$, by a certain distance, $\delta$. We name this approach ``neighbor approximation". This approximation is based on the fact that the distance from the tape where the magnetic field is influenced by the non-uniformity of $J$ is of the order of the width of the tape (figure \ref{f.distortion}). In more detail, in order to obtain $J$ (or other state variable) in turn $i$, the numerical method (in our case MMEV) calculates $J$ in the set of turns $T_i$ that follow $|{\bf r}_{j}-{\bf r}_{i}|\le \delta$, where $j\in T_i$, and ${\bf r}_{j}$ and ${\bf r}_{i}$ are the central positions of the cross-section of turns $j$ and $i$, respectively. In the calculation, it is fixed a certain total current in each turn of $T_i$ and the applied magnetic field ${\bf B}_a$ and vector potential $A_a$ created by the rest of turns, where $J$ is assumed uniform. The values of ${\bf B}_a$ and $A_a$ are
\begin{eqnarray}
{\bf B}_a({\bf r})=I\sum_{j\notin T_i} \frac{1}{S_j}{\bf b}_j({\bf r}), \label{Banei}  \\
A_a({\bf r})=I\sum_{j\notin T_i} \frac{1}{S_j}a_j({\bf r}). \label{Aanei}
\end{eqnarray}
Note that $J$ is calculated in all turns of $T_i$, not only in $i$. Thus, shielding currents in the surrounding turns or the turn of study are taken into account. The computing time for the whole coil without any approximation follows roughly $t=KN^2$ \cite{pancaketheo}, where $N$ is the number of turns and $K$ is a constant. If there are $n_T$ turns in all $T_i$, the computing time follows $t=Kn_T^2N$. Then, if $n_T^2<N$ the neighbour approximation reduces the computing time. Extra time reduction can be achieved if only one (or few) turns $i$ are calculated.

For stacks of pancake coils\footnote{Solenoids can be approximated as a stack of pancake coils.}, it may be convenient to take each pancake coil as an indivisible unity. That is, $J$ is either assumed uniform in all the turns of the pancake coil or the numerical routine calculates the detailed $J$ in the whole pancake. Then, $T_i$ contains all the turns where the turn of study $i$ is calculated and its neighbouring pancakes up to a certain order, for instance 1st or 2nd neighbors. Note that for this case, it is enough to compute $J$ for any turn $i$ of a certain pancake in order to obtain $J$ for all its turns, since any $i$ in the same pancake results in the same set of turns $T_i$. The convenience of this approach is clear for coils that their radial thickness is smaller than the tape width. For this case, the neighbour approximation reduces the computing time if $n_{\rm nei}^2<n_p$, where $n_{\rm nei}$ is the number of pancakes in $T_i$ and $n_p$ are the total number of pancakes. Further time reduction is obtained when only the AC loss in few pancakes are calculated, for instance the end pancakes.

For closely packed pancake coils with a thickness larger than the tape width, it is also useful to keep the pancake coils as an indivisible unity and assume the continuous approximation \cite{prigozhin11SST} for each pancake coil. This approximation substitutes the coil cross-section by a continuous object with effective critical current density $J_{c,{\rm eff}}={nI_c}/{S_{\rm pan}}$, where $n$ is the number of turns in one pancake and $S_{\rm pan}$ is the cross-section area of the whole pancake, including the separation between turns. The current constrain is defined by $\int J\dif z={nI}/{D}$,
where the integral is done across the tape width and $D$ is the total pancake thickness in the radial direction. In MMEV, this constrain is set directly by approximating the real pancake by another one with lower number of turns ($n_{\rm eff}$), no separation between turns, the same radial thickness, and effective transport current $I_{\rm eff}=nI/n_{\rm eff}$. For other methods, it may be necessary to use Lagrange multipliers and solve the scalar potential \cite{prigozhin11SST,zermeno13prp}.

In principle, the neighbour approximation is also applicable to coils made of conductor with magnetic parts. The reason is that the distortion of the magnetic flux density due to $J$ non-uniformity will still expand at distances of the order of the tape width, also after taking into account the interaction with the magnetic parts.


\section{Studied coils}
\label{s.coils}

In this article, we analyze two examples: one taking the real geometry of the pancake coils (real thickness of the superconducting layer and spacing between turns) and the other one taking the continuous approximation for each pancake coil. We name these coils ``detailed" and ``continuous", respectively. Their number of turns and assumptions for $J_c$ are outlined in table \ref{t.coils}. For both coils, the width and thickness of the tape are 3.96 mm and 1.4 $\mu$m, respectively\footnote{With the term `tape' we mean the superconducting layer of the coated conductor, not the whole coated conductor.}, the separation between turns is 188 and 465 $\mu$m in the radial and vertical directions, respectively, and the inner radius is 29.5 mm. For the ``detailed" coil we take the anisotropy and magnetic field dependence of $J_c$ from \cite{pardo12SSTb}, related to a particular SuperPower tape \cite{SuperPower} at 77 K. For simplicity, we assume a constant $J_c$ for the ``continuous" coil corresponding to 100 A of critical current. For the `continuous' coil, $I_c=100$ A requires temperatures below 77 K. In particular, $I_c$ measurements at low temperatures for recent tapes suggest that this $I_c$ at magnetic fields up to 6 T is achieved at a certain temperature between 50 and 65 K (see figures 1 and 2 of \cite{selvamanickam12SST}). These temperatures could be obtained with cryocooler-based systems, either by conduction cooling \cite{pooke09IES} or forced convection with helium gas. At present, temperatures down to 20 K are being used in certain HTS magnets \cite{pooke09IES}. The maximum magnetic field at the bore center is 0.3 and 4.7 T for the ``detailed" and ``continuous" coils, respectively. 

This work focuses on coated conductors with non-magnetic substrate. At present, commercial IBAD-based coated conductors use non-magnetic substrates \cite{SuperPower,SuNAM}, as well as certain RABiTS-based prototypes \cite{rupich13IES}. For tapes with magnetic substrate, several works have shown that the loss contribution of the substrate and the loss increase of the superconducting layer that it causes are important in tapes \cite{EUCAS10fmsc}, single or double pancake coils \cite{ainslie11SST,zhangM12APL} and, in principle, also solenoids with few layers. However, in packed stacks of pancake coils like in magnets, the local magnetic field is much larger, saturating the magnetic materials and minimizing their influence. Thus, models neglecting the magnetic substrate should provide a good AC loss estimate for magnet-like windings, also when the tape contains magnetic parts.

The computations use 1 current element in the tape thickness for the ``detailed" geometry and 20 elements in the radial direction of the pancakes for the ``continuous" case, between 76 and 154 elements across the tape width (the highest value is for the lowest current amplitudes, $I_m$), and a tolerance of $J_c$ between 0.20 and 0.035 \% of $J_c(B=0)$ (the lowest values are for the lowest $I_m$).

\begin{table}
\caption{\label{t.coils}The studied coils. The continuous approximation \cite{prigozhin11SST} is taken for the ``continuous" coil. More data in the text.}
\small\rm
\vspace{3mm}
\begin{tabular}{llll}
Coil name & Turns & Pancakes & Assumption \\
& per pancake & & for $J_c$ \\
\hline
``detailed" & 24 & 32 & Dependent on $B$  \\
& & & and its orientation. \\
``continuous" & 200 & 32 & Constant
\end{tabular}
\end{table}


\section{Results and Discussion}

This section presents the results of the current density and AC loss for the two studied coils: the ``detailed" and ``continuous" coils (see table \ref{t.coils} and section \ref{s.coils} for more details). The applicability of the ``uniform" and ``neighbour" approximations are also discussed.

\subsection{``Detailed" coil}

\begin{figure}
\centering
\includegraphics[width=11cm]{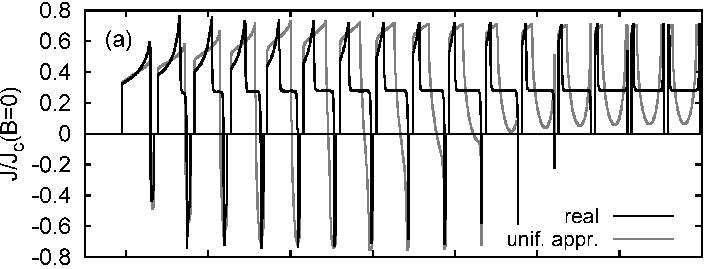}\\%
\includegraphics[width=11cm]{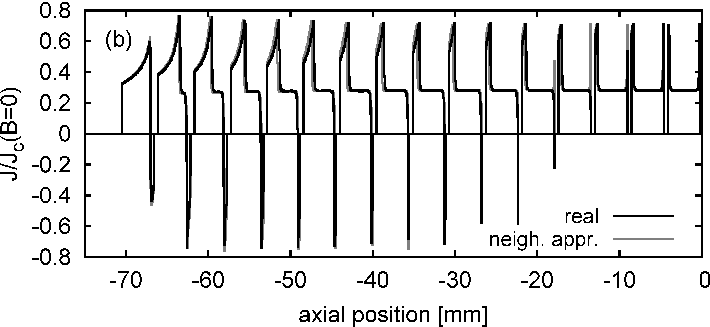}%
\caption{\label{f.Jcen} Average current density across the tape thickness for the 12$^{th}$ turn starting from the inner radius for the ``detailed" coil (table \ref{t.coils}). The uniform approximation (unif. appr.) fails to predict the current density (a), while the neighbor approximation (neigh. appr.) is satisfactory (b). The calculated case is for the peak of the AC current with 46 A amplitude. Only the lower half of the coil is shown, although the whole coil is calculated.}
\end{figure}

The current density in the ``detailed" coil presents the following features (figure \ref{f.Jcen} shows the average current density across the superconductor thickness for the central turn of each pancake: 12$^{th}$ turn from the inner radius). The uniform approximation fails to predict the current distribution (figure \ref{f.Jcen}a). The reason is that in the real case, the neighboring turns in the radial direction shield the magnetic field created by the rest of the turns, while the uniform approximation neglects this effect. As a result, for the uniform approximation the magnetization currents are larger and the plateau at the turn center disappears. In addition, the actual current distribution in the neighboring pancakes influences the magnetic field. In the real case, the neighboring pancakes partially expel the magnetic field, which concentrates between the pancakes. This increases the magnetic field close to the tape edges, locally reducing $J_c$. As a consequence, $J_c$ decreases compared to $J_c$ for the uniform approximation (see figure \ref{f.Jcen}a). On the contrary, taking into account the neighbor approximation of 1st order already produces a satisfactory result (figure \ref{f.Jcen}b). The approximation improves with approaching to the coil ends. The cause is that the average distance from the rest of the superconducting turns is the largest, and thence the magnetic field distortion from the $J$ non-uniformity is the lowest.

\begin{figure}[tbp]
\centering
\includegraphics[width=11cm]{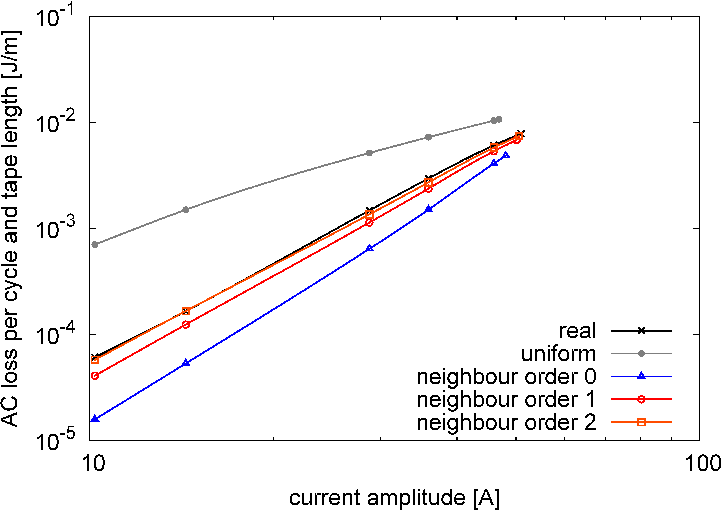}%
\caption{\label{f.Q24x32} The AC loss for the ``detailed" coil shows that the results for the ``neighbor" approximation converge to those for the ``real" geometry with increasing the order of the neighbors involved, while the ``uniform" approximation introduces a large error.}
\end{figure}

The validity of the uniform and neighbor approximations can be quantified by the error in the total AC loss (figure \ref{f.Q24x32}). The uniform approximation largely over-estimates the AC loss because it over-considers the magnetization currents. The difference increases with decreasing the current amplitude. The coil rated current (critical current of the weakest turn) is computed to be 51 A, compared to 128 A for an isolated tape. Again, the neighbor approximation provides a satisfactory result. The obtained values are better than for the uniform approximation, also for the neighbor approximation of 0th order. That is, $J$ is assumed uniform in the tapes in all pancakes, except the one where $J$ is calculated. The neighbor approximation slightly under-estimates the AC loss. The error increases with decreasing the current amplitude because at low amplitudes the current concentrates close to the tape edges, and thence it is less uniform. 


\subsection{``Continuous" coil}

The results for the ``continuous" coil qualitatively describe real-size magnets. These results are the following.

\begin{figure}[tbp]
\centering
\includegraphics[width=11cm]{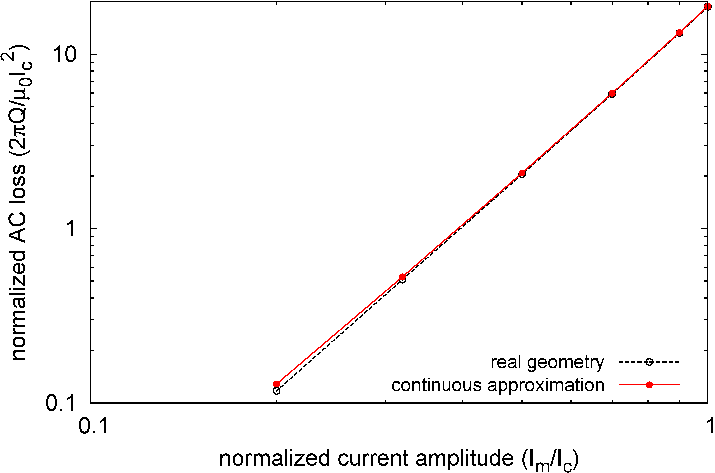}%
\caption{\label{f.Qcont1}Normalized AC loss ($2\pi Q/\mu_0I_c^2$, where $Q$ is the AC loss per cycle and tape length and $I_c$ is the critical current) for one of the 200-turn pancake coils composing the ``continuous" coil (table \ref{t.coils}).}
\end{figure}

First, we present the AC loss for one isolated pancake with the same parameters as those composing the ``continuous" coil (figure \ref{f.Qcont1}). For this case, we use the continuous approximation (section \ref{s.methlarge}). As expected \cite{prigozhin11SST}, the continuous approximation practically does not introduce any error, except a slight disagreement at low current amplitudes. Then, the error due to the continuous approximation is expected to be also negligible for the full ``continuous" coil.

\begin{figure}[tbp]
\centering
\includegraphics[width=11cm]{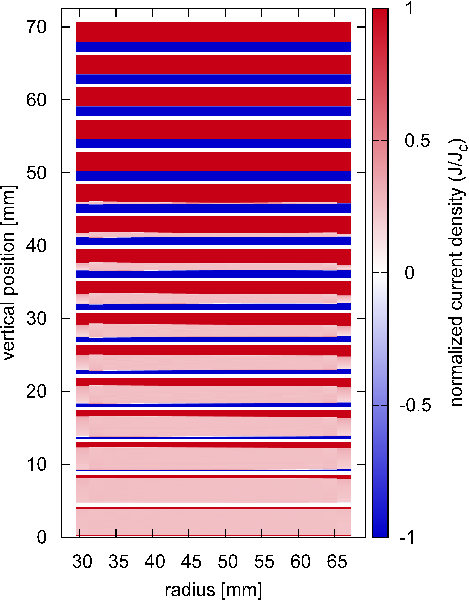}%
\caption{\label{f.Jcen200x32} Upper half of the cross-section of the ``continuous" coil (the model calculates the whole coil). The current density is for the peak of the AC cycle with amplitude $0.32I_c$. }
\end{figure}

The current density for the full ``continuous" coil by assuming only the continuous approximation is in figure \ref{f.Jcen200x32}. The five top pancakes are saturated with magnetization currents, while in the other pancakes there appears a sub-critical region with roughly uniform current density. For the unsaturated pancakes, the critical region is wider at the mid radius, its width decreases with approaching to the inner and outer radius and sharply increases at those radius. The causes are that, first, the radial magnetic field presents a maximum close to the mid radius and, second, the magnetic shielding effect from the magnetization currents drops at both edge radius (see figure 4 in \cite{pardo03PRB}).

\begin{figure}[tbp]
\centering
\includegraphics[width=11cm]{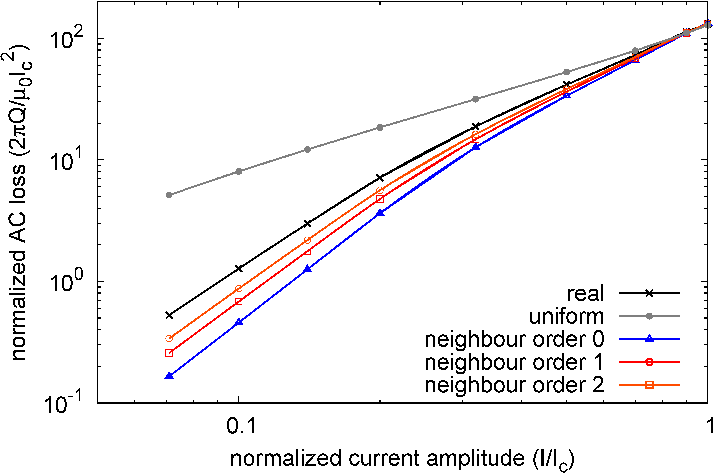}%
\caption{\label{f.Q200x32}The same as figure \ref{f.Q24x32} but for the ``continuous" coil (the ``real" case takes only the continuous approximation\cite{prigozhin11SST}). The AC loss normalization is the same as in figure \ref{f.Qcont1}.}
\end{figure}

The AC loss in the ``continuous" coil without making further approximations presents a change in slope, from around 2.6 at low current amplitudes to around 1.7 at high amplitudes (see figure \ref{f.Q200x32}). The current amplitude where this change of slope appears corresponds to the saturation by magnetization currents in the end pancakes, which contribute the most to the AC loss. Therefore, this change of slope manifests that the magnetization loss is dominant.

One may expect that the uniform approximation is accurate for magnets with many turns because in the turns that contribute the most to the AC loss (top and bottom pancakes \cite{pardo12SSTb}), the local magnetic field is much larger than the self-field of one turn (for the situation in figure \ref{f.Jcen200x32}, in the central turn of the top pancake the section average self-field is around 10 mT, while the radial component of the background magnetic field is around 530 mT). However, the uniform approximation largely over-estimates the AC loss except close to the critical current (see figure \ref{f.Q200x32}). The reason is that the uniform approximation neglects the shielding currents from the whole pancake, which are much larger than for one single tape. These shielding currents decrease the local radial magnetic field in the top pancake from around 530 mT for uniform currents to 280 mT in the real case (at the boundary between $J=+J_c$ and $J=-J_c$, figure \ref{f.Jcen200x32}). The effect of the shielding currents decreases with increasing the transport current amplitude. The cause is that with increasing the net current both the magnetic field created by the other pancakes increases and the shielding currents decrease.

The error caused by taking the neighbor approximation is low, decreasing with increasing the order of the neighbors (figure \ref{f.Q200x32}). However, this approximation is less effective for the ``continuous" coil as for the ``detailed" one (compare figures \ref{f.Q24x32} and \ref{f.Q200x32}) because the pancakes for the former are one order of magnitude wider than for the latter, and thence the magnetic field due to the non-uniform current density expands to larger distances.


\section{Summary and conclusions}

Summarizing, in this article we have discussed the current density and AC loss in coils with a high number of turns, in particular two coils with 768 and 6400 turns, respectively. We have also presented the neighbor approximation and we have found that the error that this approximation introduces is small. However, the usual uniform approximation ($J$ assumed uniform except in the turn of study) fails to predict the current density and the AC loss. The reason is that the magnetic field generated by the magnetization currents is comparable to the background magnetic field, also for coils with thousands of turns.

In conclusion, we have found that for coils made of stacks of pancakes with closely packed thin conductors, it is necessary to use relatively detailed calculations in order to achieve a satisfactory AC loss estimate. Combining the neighbor approximation with the continuous approximation, it is feasible to accurately calculate the AC loss in coils with virtually any number of turns.


\section*{Acknowledgements}

The author acknowledges M. Staines for discussions. The research leading to these results has received funding from the European Union Seventh Framework Programme [FP7/2007-2013] under grant agreement no. NMP-LA-2012-280432 and Euratom FU-CT-2007-00051.

\appendix

\section{Magnetic field and vector potential generated by a thin circular loop}

This appendix outlines the methods to calculate factors $a_j$ in equations (\ref{Aungen}) and (\ref{Aanei}) for circular coils (we assume cylindrical symmetry). These factors $a_j$ are the vector potential in Coulomb's gauge created by turn $j$ per unit current $J$, assuming uniform $J$.

The main calculation technique for $a_j$ is the following. First, all turns are divided into elements in the same way, both those where the actual current density is computed and those where $J$ is assumed uniform. For simplicity, we use uniform mesh with rectangular elements. Next, the mutual inductance between all elements is calculated following the method in the appendix of \cite{pancaketheo}. Afterwards, we use the relation between the average vector potential in the volume of element $k$, $A_k$, and the mutual inductance between elements $k$ and any other element $l$, $M_{kl}$,
\begin{equation}
\label{AkMkl}
A_k=\frac{1}{2\pi r_k}\sum_l I_lM_{kl},
\end{equation}
where $r_k$ is the mid radius of element $k$ and $I_l$ is the current in element $l$. The deduction of this relation is the following. The starting point is that $M_{kl}=U_{kl}/I_kI_l$, where $U_{kl}$ is the interaction energy between circuits $k$ and $l$ and $I_k$, and $I_l$ is the current in them. Then,
\begin{equation}
M_{kl}=\frac{1}{I_kI_l}\int_{V_k} J_kA_l \dif V \approx \frac{J_k}{I_kI_l} \int_{V_k} A_l\dif V=\frac{1}{S_kI_l}\int_{V_k}A_l\dif V.
\end{equation}
In this equation, $V_k$ and $S_k$ are the volume and cross-section surface of the elements, respectively, $\dif V$ is the volume differential, $J_k$ is the current density in element $k$, and after the approximation symbol we have taken that $J_k$ is uniform in the element. On the other side,
\begin{eqnarray}
A_k & = & \frac{1}{V_k}\int_{V_k}A\dif V = \frac{1}{V_k} \sum_l \int_{V_k} A_l \dif V = \frac{S_k}{V_k} \sum_l I_lM_{kl} \nonumber \\
& = & \frac{1}{2\pi r_k} \sum_l I_lM_{kl},
\end{eqnarray}
recovering (\ref{AkMkl}). After the last equal symbol, we have used that for rectangular elements (or any other element cross-section with radial width much lower than the average radius) $V_k=2\pi r_k S_k$. Finally, we obtain $a_j$ using (\ref{AkMkl}). Actually, we calculate the average $a_j$ on any element $k$, $a_{j,k}$, in the turns where $J$ is numerically calculated
\begin{equation}
a_{j,k}=\frac{A_k}{J_j} = \frac{1}{2\pi r_k J_j} \sum_{l\in C_j} I_l M_{kl} = \frac{1}{2\pi r_k} \sum_{l\in C_j} S_l M_{kl},
\end{equation}
where $C_j$ is the set of elements belonging to turn $j$ and at the last step we used that $I_l=J_j/S_l$. We applied the method above to calculate the neighbour approximation. The advantage is that it is very fast, once the mutual inductances are calculated. Then, it is ideal for comparison with the real geometry, where the mutual inductances need to be calculated anyway. However, there could be problems of computing time and memory for very large number of turns.

The following method, which does not require computing mutual inductances, has been applied for the uniform approximation. The term $a_j$ can be calculated as 
\begin{equation}
\label{Auncyl}
a_j=\int_{S_j}\dif r' \dif z' a_l(r,r',z-z'),
\end{equation}
where $a_l(r,r_l,z)$ is the vector potential created by an infinitely thin circular loop of radius $r_l$ concentric with the cylindrical coordinate system ($r$ and $z$) and located at $z=0$. The expression of $a_l(r,r_l,z)$ is \cite{landau}
\begin{equation}
a_l=\frac{1}{\pi k}\sqrt{\frac{r_l}{r}}[F(k)(1-k^2/2)-E(k)]
\end{equation}
with
\begin{equation}
k=\sqrt{\frac{4r}{(r_l+r)^2+z^2}},
\end{equation}
where $F(k)$ and $E(k)$ are the complete elliptic integrals of the first and second kind, respectively. This integral can be calculated numerically by dividing the turn into elements, where it is assumed that the current is concentrated at the cross-sectional center. Then,
\begin{equation}
a_j(r,z)\approx \sum_{k\in C_j} a_l(r,r_k,z-z_k) \Delta r \Delta z,
\end{equation}
where $z_k$ is the central axial position of element $k$ and $\Delta r$ and $\Delta z$ are the radial and axial dimensions of the elements, respectively. Note that for our case, the coordinates ($r$,$z$) are always outside turn $j$. Given a number of elements in the turn wider direction, $m$, we divide the tape turn in $m$ elements in the wider direction and as many elements in the other direction to obtain a cross-sectional aspect ratio of the elements as close as possible to 1. In order to calculate $a_j$ up to a desired tolerance, we increase progressively the number of elements, as follows. We set $m=1$ as initial value and calculate $a_j$, duplicate $m$ and calculate $a_j$ again, and repeat the process until the relative difference between the previous value is below a certain tolerance. Once the procedure to calculate $a_j$ is settled down, we calculate the average $a_j$ in any element $k$, $a_k$, in the turns where $J$ is calculated. This can be done by dividing element $k$ into sub-elements
\begin{equation}
a_{j,k} \approx \frac{1}{S_k} \sum_{s\in C_k} a_j(r_s,z_s) \delta r \delta z.
\end{equation}
In this equation, $C_k$ is the set of sub-elements dividing element $k$ and $\delta r$ and $\delta z$ are the radial and axial dimensions of the sub-elements, respectively. In order to calculate $a_{j,k}$ up to a certain tolerance, we use the same method as for $a_j$, described above. In this article, we use a tolerance of 0.005 \% for both $a_j$ and $a_{j,k}$, resulting in a combined tolerance of 0.01 \%. Test simulations showed that reducing the tolerance below this value does not appreciably improve the results.

Both methods to calculate $a_j$ have been compared for coils with few turns, presenting identical results within the tolerance.

\section*{References}

\end{document}